\documentclass[a4paper]{article}
\usepackage{amsmath}
\usepackage{amssymb}
\usepackage{graphicx}
\usepackage{subfig}

\begin{document}

\title{Digital signatures with Quantum Candies}
\date{\today}
\author{Tal Mor \and Roman Shapira \and Guy Shemesh}

\maketitle
\setcounter{footnote}{0} 

\begin{abstract}
Quantum candies (qandies) is a pedagogical simple model which describes many concepts from quantum information processing (QIP) intuitively, 
without the need to understand or make use of superpositions, and 
without the need of using complex algebra.
One of the topics in quantum cryptography which gains research attention in recent years is quantum digital signatures (QDS), involving protocols to securely sign classical bits using quantum methods.

In this paper we show how the ``qandy model'' can be used to describe 
three QDS protocols, 
in order to provide an important and potentially practical 
example of the power of 
``superpositionless'' quantum information processing, 
for individuals without background knowledge in the field.
\end{abstract}

\section{Introduction}

Quantum information processing (QIP) is an intensively studied field in the academia, ever since Feynman had originally proposed the idea of quantum computing in 1982 \cite{feyn82} and to this day.
In the recent decade, even many technology companies in the industry (including
Google, IBM, Intel, Microsoft, Alibaba and others) 
direct more and more resources towards quantum computing and communication research and development.

Nowadays, one of the most prominent research directions in QIP (and specifically in quantum cryptography) is the topic of quantum digital signatures (QDS).
Before we discuss QDS, we briefly mention the historical background and motivation to QDS - classical digital signatures.
The idea of digital signatures was first proposed by Diffie and Hellman \cite{dh76} in 1976.
The goal is to allow Alice to digitally sign a message for Bob, such that Alice cannot later repudiate signing the message and Bob (or any other malevolent party) cannot forge Alice's signature.
Since then, many classical digital signature schemes have been proposed, including the famous Lamport's one-time digital signature scheme \cite{lam79} (which we will further discuss in Section~\ref{sec:ds}).
The main downside of typical classical signature schemes is that their security relies on unproven computational assumptions, such as the hardness of factoring large numbers.
With the advent of Shor's algorithm \cite{shor99}, which allows factoring numbers efficiently on a quantum computer, a need arises to come up with digital signature protocols whose security doesn't rely on such computational assumptions.

In 2001, Gottesman and Chuang \cite{gc01} have first proposed the notion of ``Quantum Digital Signatures''.
Their general idea was to use qubits instead of classical bits to sign the message, thus potentially achieving information-theoretically secure signature protocols, similar to QKD protocols like \cite{bb84}.
However, Gottesman and Chuang's proposed scheme serves more as a theoretical base model for QDS schemes and is quite unfeasible to realize physically, due to several major shortcomings:
1. The need for costly quantum comparison operations (``SWAP'' tests) between distant parties,
2. The need for long-term reliable quantum memory,
3. The need for authenticated quantum channels between the parties (namely, quantum states sent by the sender are received fully identically by the recipient).

Since Gottesman and Chuang's work, much research has been done in attempts to overcome said shortcomings.
In 2006, \cite{acj06} came up with a novel coherent state comparison method, as well as a QDS protocol based on this method.
The main advantage was that the coherent state comparison is more feasible to physically implement than the SWAP test, as it requires only linear optics and photon detectors
(which are used to construct a ``multiport'' comparison device).
However, the \cite{acj06} scheme still required the usage of quantum memory.
Then, in 2014, \cite{dwa14} improved the \cite{acj06} scheme and proposed a QDS that still required a ``multiport'' device, yet without the requirement of quantum memory.

Most relevant for our work, in \cite{wdka15}, the authors have proposed several QDS protocols relying on QKD components, allowing for highly realizable protocols without the need for optical multi-ports or quantum memory.
One of the protocols in \cite{wdka15} is a ``classical'' signature protocol that
relies on secret shared keys, which may be distributed by QKD methods (protocol
P2).
Lastly, the authors in \cite{awka16} took the ``classical'' protocol from \cite{wdka15} one step further and proposed a QDS protocol without the need for authenticated quantum channels.
Thus the protocol of \cite{awka16} may be considered as the modern day ``standard'' feasible QDS scheme.
See also \cite{aa15} for a detailed review of existing QDS protocols.

Although quantum information processing gains a lot of popularity, understanding its fundamental rules requires familiarity with non-trivial concepts from physics and mathematics, such as superpositions, unitary transforms, and complex linear algebra in general.
In an attempt to improve accessibility to this important field of science, an idea for a simplified model was first proposed by Jacobs --- the ``Quantum Candy'' model.
The model was later greatly expanded and defined in \cite{lm20,lms21}.
In essence, the ``Qandy Model'' abandons the notion of quantum bits (qubits) in
favor of ``Quantum Candies'' (qandies) - mystical candies which similarly to
classical bits only have a discrete set of possible states, but unlike classical
bits do behave essentially ``quantumly''.
It is shown and discussed in \cite{lm20,lms21}, that many concepts from QIP, and especially quantum cryptographic protocols, can be interpreted and defined in the qandy model.
However, not many quantum-information protocols can be defined without using superpositions, unitary transformations, and phases.
In other words, a limited, ``superpositionless'' variant of quantum theory may be useful for some non-classical features yet not for others.

In this paper, our goal is to present quantum digital signature protocols with QKD components, relying on \cite{wdka15, awka16}, using qandies instead of qubits.
We discuss three qandy signature protocols, and provide basic intuition for their security.
Detailed security analysis is left for future research.
The paper is structured as follows: In Section~\ref{sec:ds} we discuss computationally secure classical digital signatures.
In Section~\ref{Sec:OTPS} we present the classical digital signature variant (due to~\cite{wdka15}) of the P2 protocol of \cite{wdka15}, which is an unconditionally secure signature scheme.
In Section~\ref{sec:qandies} we discuss the qandy model and its basic rules.
In Section~\ref{sec:qandy_ds} we present the first proposed qandy digital signature protocol, based on protocol P1 of \cite{wdka15}, and provide some remarks and discuss its security.
In Section~\ref{sec:alt_sig} we present and discuss subsequent improvements of the first protocol, namely P2 of \cite{wdka15} and the protocol of \cite{awka16}, and argue they may be implemented using qandies as well.
We finish the paper with our conclusions in Section~\ref{sec:conclusion}.

\section{Computationally Secure Digital Signatures} \label{sec:ds}

We begin this section by introducing the goals of Digital Signatures (DS) schemes, their usages today and present a definition.
We will then shortly describe the Lamport's signature scheme \cite{lam79} - a simple classical scheme that is based on the concept of one-way functions (OWF).

\subsection{Digital Signatures - Motivation} \label{subsec:motivation}
Let us examine the following scenario: a sender, Alice, sends the message $M$ to the recipient Bob.
Is there a way for Bob to ensure that Alice is indeed the sender of $M$?
In other words, can Bob find a way to ensure that a malicious third party, Eve, was not the one sending $M$ in the name of Alice, or tempered with its content?

This type of scenario happens all the time in our era of digital communications.
For example, during credit card transactions, a bank getting a request from one of its clients to transfer money to another account, needs to make sure the request was indeed made by the client.
We will call this desired property \textbf{unforgeability}. 

What happens if Alice made a future promise to Bob through a message $M$ (and signed it), and later on denies she is the one who signed it?
Namely, she claims someone forged her message.
In this case Alice is the dishonest party, and we would like to prevent her from successfully executing such trickery.
A digital signature scheme preventing this behavior is said to promise \textbf{non-repudiation}.

Lastly, a good digital signature scheme should promise a recipient who accepted a signature that any other recipient would accept it as well.
This property is called \textbf{transferability}.
Transferability  is closely related to the second property of non-repudiation, and many times these two properties are identical - for example when a dispute is settled by majority decision.
One can also think of bad schemes when the two are not entirely equivalent, for example when transferability is not promised even when Alice is honest.


\subsection{Definition and the Lamport Scheme} \label{sec:Lamport}

The classical digital signature scheme has a conventional, 
formal definition (the interested readers may
see the formal approach in Appendix~\ref{sec:append}).
The definition promises that every scheme that meets its requirements, will have the three properties mentioned in Subsection~\ref{subsec:motivation}: unforgeability, non-repudiation and transferability.
For clarity and simplicity, in this paper we only deal with the following setup: 
There are three parties, the messages are all one-bit message (0 or 1) and there is at most one dishonest party.
We will now present the ``canonical'' example for a secure one-time DS scheme - the Lamport scheme.
Instead of proving its security by the formal definition, 
we will refer to the unforgeability and non-repudiation properties.
We will not refer to transferability since in the Lamport scheme (and more generally, whenever a dispute between three parties is settled by majority vote) it is equivalent to non-repudiation.

\paragraph{The Lamport Scheme}
The Lamport scheme \cite{lam79} is a one-time signature scheme based on the usage of a one-way function. Intuitively, a one-way function is a
function that is easy to calculate but hard to invert, i.e.,
for any input $x$ one can efficiently calculate $H(x)$,
but it is hard to find $x'$ such that $H(x')=H(x)$, given only $H(x)$.
The Lamport scheme can be viewed as a parallel to the one-time pad (OTP) scheme used in symmetric key encryption, for it is arguably the simplest application of a DS scheme given a OWF.
There is however an important difference between the two: the OTP scheme does not require \textbf{any} assumptions for it to be information-theoretically secure, while the Lamport scheme uses a one-way function. 
The existence of such functions has yet to be proven (though there are some candidates, see \cite{gol01}), which makes the Lamport scheme only \emph{possibly} computationally secure.

We will present a version of the Lamport scheme for one-bit messages. 
We will also repeat this explanation in a formal way, and using 
formal definitions in Appendix~\ref{sec:append}.

\paragraph{}
Let Alice be the sender of a message $m \in \{0,1\} $, Bob the recipient and Charlie the arbiter in the case of dispute between Bob and Alice. Let $H: \{0,1\}^n\to\{0,1\}^n$ be a one-way function.

\begin{enumerate}
    \item \textbf{Private key generation}
    For each value of $b\in \{0,1\}$, Alice generates a random string $X_b\in\{0,1\}^n$.
    Both $X_b$ serve as the private keys of Alice.

    \item \textbf{Public key distribution}
    For each value of $b$, Alice calculates and publicly announces $P_b=H(X_b)$.
    Both $P_b$ serve as the public keys. 

    \item \textbf{Signed messaging}
    If Alice wishes to send Bob a signed one-bit message $m$, she sends $(m, X_m)$. 
    I.e, the signature of the message $m$ is\footnote{Note that in this construction, $sk=[\sigma_0,\sigma_1]$} $\sigma_m=X_m$

    \item \textbf{Verification}
    When receiving $(m, X_m)$ from Alice, Bob (who knows $H$ and $P_m$) accepts the message if and only if $H(X_m)=P_m$.

    \item \textbf{Arbitration}
    A dispute occurs when Bob claims Alice sent a message $m$, and Alice denies it.
    If Bob is dishonest, this is the forgery attempt case.
    If Alice is dishonest, this is the repudiation attempt case.
    A dispute is settled in the following way: Bob sends Charlie the value of $X_m$ he claims he received from Alice, $X^B_m$.
    Charlie then checks if $H(X^B_m)=P_m$, and if so he declares Alice as the dishonest one.
    If not - he declares Bob as the dishonest one.

\end{enumerate}

\textbf{Unforgeability} is guaranteed since if $H$ is a one-way function, Bob will not be able to find such $X^B_m$ (within reasonable time).
\textbf{Non-repudiation} is guaranteed for the same reason - if Bob managed to present such $X^B_m$, then he definitely got it from Alice.

\section{One-Time Pad Signatures} \label{Sec:OTPS}

One-time pad (OTP) is the popular name for the one-time pad cipher
also known as the Vernam cipher or the Vernam code.  There is some ambiguity
regarding those terms so we will clarify here precisely what we mean:
Vernam cipher or OTP cipher is the use of a random string of bits pre-shared
by a sender Alice and a receiver Bob, named the ``pad'',
to send secret strings (of length equal to the length of the pad), by
using the bit-wise XOR operation between bits of the pad and bits of the 
secret message. It is crucial for the security of the transmitted secret that
the pad is only used once, and hence the term one-time pad.

We refer here to the random and pre-shared pad as the OTP, and we present
a signature algorithm invented by~\cite{wdka15}, which we name one-time pad
signature (OTP-S). The paper~\cite{wdka15} relies, in their protocol P2,
on quantum key distribution, for generating the OTP itself.  But we can ignore,
for now of course, the quantum portion of the protocol, if we assume that the OTP
is already shared between each pair of users.

The OTP-S of \cite{wdka15} is not the first unconditionally-secure signature scheme, and earlier ones were suggested (\cite{hsz+00, ss11, cr90}).
Here is a quote from the Discussion section in~\cite{wdka15}:
``P2 differs because it is an information-theoretic-secure classical digital signature scheme relying only on secret shared keys, without further assumptions such as a trusted third party or authenticated broadcast channels \cite{hsz+00, ss11, cr90}. 
This illustrates how novel classical protocols can arise inspired by quantum information science.''

\subsection{Protocol Assumptions}

As before, we assume three parties, Alice, Bob and Charlie,
and arbitration is required if 
Bob provides a document (one bit in our case) claiming it is signed by Alice,
and Alice said she did not sign this document.
We assume that in this protocol (and all subsequent protocols) the arbiter Charlie is always honest.

Let us assume that each pair of users share in advance 
random strings (secret classical keys --- pads) of any desired length, 
and that each secret random bit is used just once, hence it is a one-time 
pad.
But it is not a one-time pad cipher.

We note that a small number of bits from the pre-shared secret strings can be used for authentication via universal hash functions \cite{cw79, wc81}.
Thus we assume that between each pair of users there also exists an authenticated classical channel, and that each transmission (even via OTP) is authenticated.

\subsection{Protocol Specification} \label{subsec:otps-spec}
We now outline the classical protocol; it is nearly precisely P2 of 
Subsection~\ref{sec:P2}, but without the quantum parts:
\begin{enumerate}
    \item 
    \textbf{Private Key Generation} For each future message bit $b\in \{0, 1\}$, Alice generates two random strings $X^{B}_{b},X^{C}_{b}\in\{0,1\}^n$ (Total of 4 strings).
    
    \item 
    \textbf{``Public'' Key Distribution} For each future message bit $b\in \{0, 1\}$, Alice sends $X^B_b$ to Bob and $X^C_b$ to Charlie, using OTP.
    
    \item \label{step:pk_symm_OTP} 
    \textbf{``Public'' Key Symmetrization} For each future message bit $b \in \{0, 1\}$, each of Bob and Charlie decides on a random subset of $n/2$ bits from $X^B_b$ ($X^C_b$) and forwards them (value and index) to the other, using OTP.
    Bob's decision on the forwarded subset is independent of Charlie's.
    
    \item 
    \textbf{Signed Messaging} For signing a \emph{specific} bit $b \in \{0, 1\}$, Alice sends Bob the string $(b, X^B_b, X^C_b)$ via an authenticated classical channel.

    \item 
    \textbf{Verification} Bob accepts Alice's signature if it matches his key $X^B_b$ and the indices of $X^C_b$ Charlie sent him.
    As the keys are classical, error-free and authenticated, we can assume the final shared keys must be fully identical (if the parties are honest).
    \item 
    \textbf{Arbitration} In case of a future dispute regarding the message content, Bob forwards to Charlie the string $(b, X^B_b, X^C_b)$ he obtained from Alice.
    Charlie counts the mismatches between Bob's forwarded keys $(X^B_b, X^C_b)$ versus his key $X^C_b$ and the parts of the key $X^B_b$ Bob sent him in Step~\ref{step:pk_symm_OTP}, and accepts only if:
    \begin{enumerate}
        \item There is no mismatch between Bob's forwarded $X^B_b$ and the parts of $X^B_b$ Charlie received from Bob in Step~\ref{step:pk_symm_OTP}.
        \item The number of mismatches between Bob's forwarded $X^C_b$ and Charlie's $X^C_b$ is less than $s_v \cdot n$, where $s_v$ is the arbiter's verification threshold parameter.
    \end{enumerate}
    Charlie accepting means he accepts Bob's forwarded message as valid, i.e. Alice is the dishonest party, and a rejection means Bob's forwarded message is invalid, i.e. Bob is the dishonest party.
\end{enumerate}

The protocol is proved to be secure against repudiation and forgery for a choice of $0 < s_v < 1/4$, with the full proof found in \cite{wdka15}.
Intuitively, the larger $s_v$ is, the easier it is to successfully forge (by Bob), and the lower $s_v$ is, the easier it is to successfully repudiate (by Alice).

We shall repeat similar steps when discussing P2 in Subsection~\ref{sec:P2}.
For the sake of consistency with earlier work by~\cite{wdka15,awka16},
we keep the internal order of~\cite{wdka15}, thus presenting protocol
P1 prior to protocol P2; but firstly we present quantum candies.

\section{Quantum Candies} \label{sec:qandies}
We now present the notion of quantum candies (or qandies) as presented in \cite{lm20, lms21}.
The concept of qandies was first proposed by Jacobs \cite{kayla-jacobs} --- see a detailed description of Jacobs qandies in \cite{lm20, lms21}.
The Jacobs qandies model can be viewed as a development of the ``chocolate balls'' model presented by Karl Svozil in several papers (e.g. \cite{svo06}), although the two models were independently developed.
Classically, communication between two parties is executed by transmitting pieces of information called bits, which are simply zeroes and ones.
However, in the world of quantum communication, there are transmissions of \textbf{qubits}, or quantum bits.
To create a qubit, one uses a physical system with quantum properties, such as the polarization of photons, and maps the state of the system to a qubit value.
This is similar to the classical case, when (for example) a presence of current is mapped to the bit ``1'', and absence of current is mapped to the bit ``0''.
However, due to the laws of quantum mechanics, qubits present extraordinary properties which are very different from the classical world.

Those properties have significant information-theoretic consequences (e.g. the no-cloning theorem), and the notion of quantum candies helps explaining some of those properties in a simple and mathematics free manner \cite{lm20, lms21}.
In the conference version \cite{lm20}, Jacobs qandies are extended to Lin-Mor qandies --- by adding a correlation-generating machine that generates states (of pairs of qandies) resembling entangled states in quantum theory.
Furthermore, in \cite{lm20} it is shown how a naive qandies bit commitment scheme can be cheated using such pseudo-entangled qandies.
In the much extended journal version \cite{lms21}, Lin-Mor qandies are further extended to Lin-Mor-Shapira qandies --- by adding correlation-measuring machine that can distinguish various fully-correlated states (of pairs of qandies), and furthermore presented protocols for communication and for key distributions relying on such qandies machines.

Imagine two parties, Alice and Bob, who wish to communicate with each other.
For simplicity, let Alice be the sender of a message and Bob the recipient.

We equip Alice with a machine that can generate a special kind of candies, called ``qandies''.
At any given time, each qandy has exactly \textbf{one} property out of the following options: Green color $\{G\}$, Red color $\{R\}$, Chocolate flavour $\{C\}$ and Vanilla flavour $\{V\}$.
So Alice's machine has only four buttons: {$R$, $G$, $C$, $V$}, which match the properties respectively.

When receiving a qandy, Bob can either taste it or look at it (but not both at the same time): if Alice prepared a qandy with a certain taste, Bob can taste the qandy and he will measure that taste;
e.g., if Alice pressed the \{$V$\} button and Bob decided to taste the qandy, then he will taste vanilla.
Similarly, if Alice prepared a qandy with a certain color, Bob can look at the qandy and he will see that color;
e.g., if Alice pressed the \{$G$\} button and Bob decided to look at the qandy, then he will see a green qandy.

This way Alice can send the qandies to Bob, with one color mapped to the value ``0'' and the other to the value ``1''. Similarly, one taste is mapped to ``0'' and the other to ``1''.

What happens if Alice pressed the \{$V$\} button and Bob decided to look at the candy's color?
This scenario is what makes Alice's qandies so special, and gives them their quantum-like properties that turn them into ``qandies''.
If Alice prepared a certain property (taste or color) and Bob decided to measure the \textbf{other} property, then he will measure a random outcome (out of the two possibilities for the measured property).
So if Alice pressed the \{$V$\} button and Bob decided to look at the qandy's color, he will see Red with probability of $\frac{1}{2}$ or Green with probability of $\frac{1}{2}$.

This means that only the original property prepared by Alice is meaningful, since measuring the other property leads to a totally random outcome.
To quote the original paper \cite{lm20, lms21}:
``The key feature of these qandies is that each single qandy really has only a single specific property.
This is one form of what is known as the complementarity principle in quantum physics:
if color is defined, taste cannot be defined, and
if taste is defined color cannot be defined.
The rule of complementarity applies both to the person (i.e. person/machine) preparing the qandies, and to anyone observing the qandy.''

In particular, as also shown previously in \cite{lm20} and even more so in \cite{lms21}, the qandy model (i.e. a ``superpositionless'' variant of quantum theory) is sufficiently strong for presenting interesting features of quantum protocols while being able to fully avoid the usage of superpositions or phases.
In the following sections, we will see how these characteristics of qandies allow us to successfully achieve our goal of creating secure signature schemes.

\section{A Digital Signature Protocol With Qandies} \label{sec:qandy_ds}

We now describe a digital signature protocol using quantum qandies.
The protocol is a one-time signature protocol for signing a single bit $b$, based on the P1 QDS protocol of \cite{wdka15}.
The protocol has three parties: Alice the signer, Bob the verifier and Charlie the arbiter.
We first present the protocol assumptions, then the protocol steps and lastly a general discussion of the protocol before moving on to analyze its security.

\subsection{Protocol assumptions}
\begin{enumerate}
    \item
        Alice is in possession of standard qandy generation capabilities: she can create qandies of the 4 types $\{R, G, C, V\}$.
        Alice also has an ideal coin, i.e. she can generate random classical bits; such an assumption is automatically satisfied using qandies measuring devices (e.g. generate a color qandy and measure its taste).
    \item 
        Bob and Charlie possess standard qandy measurement capabilities: they can either look at or taste the qandies.
    \item \label{itm:memory_assum}
        Bob and Charlie possess a ``qandy memory'' capability, namely they can store received qandies indefinitely and measure them later.
    \item
        Between each two parties there exists an authenticated classical channel (using a small number of pre-shared bits); no eavesdropper or noise can change the classical bits sent between the parties.
\end{enumerate}
Notes: 
\begin{itemize}
    \item 
        The assumption in item~\ref{itm:memory_assum} is not strictly necessary (the original protocol P1 in \cite{wdka15}, does not make it).
        We later discuss how the protocol may be defined without it, but we make it here to simplify the description.
    \item
        In the original protocol of \cite{wdka15}, another assumption was made: between each two parties there exists an authenticated quantum channel, i.e. no eavesdropper or noise can change the qandies sent between the parties.
        We note the implementation of an authenticated quantum channel is a far-from-trivial task \cite{bcg+02}, and it is unclear whether such an authenticated channel can be implemented using qandies.
        However, as stated by \cite{wdka15}, the requirement of authenticated
        quantum channels may be dropped if the parties use the first step of
        most standard QKD technique: the step of
        comparing measurement results (TEST), after sending quantum states to one another and measuring
        some of them (and aborting if the effective noise level, $p_e$, is too high).
        Such a technique may easily be implemented using qandies \cite{lm20, lms21}, and we thus assume no authenticated ``qandy channels'' are needed for this protocol.
    \item
        Adding TEST somewhat modifies the security analysis to make it more 
        complicated; a detailed analysis is beyond the scope of this paper, and is left
        for future work.
    \item
        We assume $n$ is very large, so that the relevant law of large numbers will work well, and any exponentially small tail can be safely neglected.
\end{itemize}

\subsection{Protocol Specification} \label{subsec:p1-spec}
\begin{enumerate}
    \item \label{step:pk_gen} 
    \textbf{Private Key Generation} For each future message bit $b \in \{0, 1\}$, Alice generates a random string (``private key'') $X_b = (x_1^b, x_2^b, \ldots, x_n^b)$ of \emph{characters} from the set $\{`R', `G', `C', `V'\}$ (each character can be represented by 2 bits).
    The key length $n$ is the security parameter of the protocol.
    \item \label{step:pk_dist} 
    \textbf{``Public'' Key Distribution} For each future message bit $b \in \{0, 1\}$, Alice creates \emph{two} copies of \emph{qandy strings} (``public keys'') $Q_b = (q_1^b, q_2^b, \ldots, q_n^b)$ using her qandy generating machine - each qandy $q_i^b$ according to the private key character $x_i^b$.
    Alice sends one copy of $Q_b$ to Bob and the other to Charlie.
    \item \label{step:pk_symm} 
    \textbf{``Public'' Key Symmetrization} For each future message bit $b \in \{0, 1\}$:
    \begin{enumerate}
        \item
        Bob (Charlie) decides on a random subset of $n/2$ qandies from his copy of $Q_b$ and forwards them to Charlie (Bob), without measuring.
        The decision on the forwarded subset is independent of the other party.
        \item \label{step:meas_res}
        Bob (Charlie) randomly looks at or tastes each of the $n/2$ qandies he has kept, as well as the $n/2$ qandies received from Charlie (Bob), and writes down the measurement results.
        \item
        Alice discloses part of the qandies to Bob and Charlie (via an authenticated classical channel), for TEST.
        There are two TESTs:
        \begin{enumerate}
            \item Between Alice and Charlie, in order to make sure Bob didn't attempt to eavesdrop in Step~\ref{step:pk_dist} (distribution).
            \item Between Alice and Bob, in order to make sure Alice didn't attempt to eavesdrop in Step~\ref{step:pk_symm} (symmetrization).
        \end{enumerate}
        In both TESTs, both parties choose which indices they want to reveal.
        Bob and Charlie abort the protocol if the noise rate $p_e$ is higher than $s_a - \delta(n)$.
        $s_a \geq \delta(n)$ is the recipient's verification threshold parameter that depends on the parameters of the qandy channel and equals 0 in the ideal case.
        $\delta(n)$ is a small parameter (function of $n$), such that the probability of a noise rate higher than $s_a$ on the unmeasured (non-TESTED) qandies is exponentially small.
        Note that in the TESTs, qandies with conjugate measurement bases (e.g. Alice sent a color Qandy but Bob tasted it) are removed from the statistics and do not contribute towards the noise rate calculation.
    \end{enumerate}
    At this stage, Bob and Charlie do not yet know the indices of the qandies that were received from the other and were not TEST qandies.
    \item 
    \textbf{Signed Messaging} For signing a \emph{specific} bit $b \in \{0, 1\}$, Alice sends Bob the string $(b, X_b)$ (via an authenticated classical channel)
    \item \label{step:ver} 
    \textbf{Verification} At this stage, Bob and Charlie disclose to one another all the indices of the qandies they forwarded in step~\ref{step:pk_symm}, via an authenticated classical channel.
    Bob counts the number of mismatched indices between his measurement results of $Q_b$ from step~\ref{step:meas_res} and Alice's candidate private key $X_b$.
    
    A mismatch for index $i$ occurs when Alice's private key character $x_i^b$ contradicts Bob's measurement result for that index: for example, $x_i^b$ is $`R'$ but Bob has a measurement that says $q_i^b$ is $G$.
    On the other hand, if $x_i^b$ is $`R'$ but Bob decided to taste and measured $C$, this is not a contradiction as he chose the wrong basis.
    
    Bob accepts Alice's signature if the number of mismatches is smaller than $s_a \cdot n$, and rejects otherwise.
    \item \label{step:arb} 
    \textbf{Arbitration} In case of a future dispute regarding the message content, Bob forwards to Charlie the string $(b, X_b)$ he obtained from Alice.
    Charlie counts the mismatches similarly to Bob, but accepts (i.e., agrees with Bob) only if the number of mismatches is less than $s_v \cdot n$, where $s_v > s_a$ is the arbiter's verification threshold parameter.
\end{enumerate}

Following are several remarks on the protocol:
\paragraph{No-Cloning of Qandies}
In step~\ref{step:pk_dist}, the requirement to create two copies of the qandy strings stems from the no-cloning theorem for qandies \cite{lm20, lms21}: Bob (Charlie) is unable to copy the qandies he obtained from Alice and send them to Charlie (Bob).
This is in contrast to the classical Lamport protocol, where Alice's public key can be copied as many times as necessary and by anyone.
On one hand, this property is a weakness of the quantum/qandy protocol compared to the classical protocol, as the number of recipients must be determined in advance.
On the other hand, this is what enables the protocol to be unconditionally secure \footnote{If we replace one honest arbiter by several such that a majority of them are assumed to be honest, the issue of potential cloning becomes problematic.}.

\paragraph{Usage of Memory}
In step~\ref{step:pk_symm}, the fact Bob and Charlie have a ``qandy memory'' allows them to select a random subset of $n/2$ qandies in advance to send to the other.
In contrast, in the P1 protocol of \cite{wdka15} Bob and Charlie decide randomly and independently for \emph{each} qandy obtained from Alice whether to measure it or forward it, with equal probabilities.
The protocol is then aborted if each of them received from the other a number of qandies too far from $n/2$ (up to a certain threshold).
Our qandy protocol could be defined analogously, removing the need for ``qandy memory''.

\paragraph{Possibilities of Elimination}
In step~\ref{step:pk_symm}, Bob and Charlie do not exchange the same random subset of qandies, since the choice is made by each of them independently.
For large $n$, after the exchange Bob and Charlie will each have zero copies of $q_i^b$ with probability $1/4$, one copy with probability $1/2$ and two copies with probability $1/4$.
This means that for an index $i$ with zero copies, they cannot eliminate any possible value of the private key and hence have to accept any claim by Alice for that index.
On the other hand, for an index $i$ with two copies, they can eliminate at most two possible values of the private key - by looking at the first copy and tasting the second.


\subsection{Security Analysis} \label{subsec:security}
Security of a digital signature protocol relies on non-forgery (by Bob) and non-repudiation (by Alice).
In the following section, we will provide a basic intuition for the qandy signature protocol being secure against forgery and repudiation, based on the full analysis of \cite{wdka15} done for quantum case.
Since we assume non-ideal/noisy ``qandy channels'' (either due to natural noise or eavesdropping activity), we discuss the probability of an ``honest abort'' (following \cite{pab+20}) and discuss the optimal choice of the parameters $s_a, s_v$.
Note that the security is information-theoretic - it does not depend on unproven computational assumptions, unlike most currently known classical protocols.

\paragraph{Security Against Forgery}
For a successful forgery, a cheating Bob has to guess a signature $(b', X_{b'})$ he did not obtain from Alice, that is consistent with the public key qandies from $Q_{b'}$ in Charlie's possession - i.e. cause less than $s_v n$ mismatches with Charlie's qandies, to pass the arbitration step.
Due to the symmetrization (step~\ref{step:pk_symm}) of the protocol, Charlie has left in his possession $n/2$ of the qandies of each public key $Q_b$.
It can be shown that Bob's best course of action is to perform a minimum-error measurement \cite{wda14} on his copy of the public key $Q_{b'}$, which in the qandy world translates to randomly looking at or tasting a qandy. If needed for step~\ref{step:pk_symm}, Bob can generate the qandy he observed and forward to Charlie.
Such strategy yields a probability of $p_f$ that Bob causes a mismatch between an honest Alice's private key element and the copy of the qandy left in Charlie's possession.

Overall, it is shown by \cite{wdka15} that for $s_v < p_f$ the probability of forgery is:
\begin{equation}
    p_{\text{forgery}} \leq \exp(-c_f(p_f-s_v)^2 n)
,\end{equation}
for some constant $c_f > 0$, thus the protocol is secure against forgery -- as the probability of forgery decays exponentially with the security parameter $n$.

\paragraph{Security Against Repudiation}
For a successful repudiation, a cheating Alice has to cause Bob to accept her signature in step~\ref{step:ver} (verification) while Charlie rejects it in step~\ref{step:arb} (arbitration).
Even if Alice sends different public keys $Q_b^{\text{Bob}} \neq Q_b^{\text{Charlie}}$ to Bob and Charlie respectively, step~\ref{step:pk_symm} (symmetrization) ensures that from her perspective, each qandy $q_b^{\text{Bob}}, q_b^{\text{Charlie}}$ has equal probability of ending up in either Bob's or Charlie's possession.
Therefore Alice does not benefit from sending different public keys to Bob and Charlie.

It is shown by \cite{wdka15} that Alice's best strategy to repudiate is to send the same public key to both Bob and Charlie (i.e. as in the honest case), but when signing, deliberately choose a signature with $n(s_v - s_a)/2$ mismatches with the public key.
Requiring $s_a < s_v$ as mentioned previously, such attempt yields a probability of repudiation:
\begin{equation}
    p_{\text{repudiation}} \leq \exp(-c_r(s_v-s_a)^2 n)
,\end{equation}
with some constant $c_r > 0$, thus the protocol is secure against repudiation.

\paragraph{Probability of an Honest Abort}
We assumed the ``qandy channel'' employed by the participants is noisy, either due to natural noise or eavesdropper activity, with probability smaller than $p_e + \delta(n)$ for a ``noise mismatch'' caused between Alice's private key element and Bob's measurement (estimated by employing TEST on some of the qandies sent in the distribution stage).
Intuitively, if Bob's verification threshold $s_a$ is too low, allowing for less mismatches, a high noise rate $p_e$ will cause Bob to abort many of Alice's (honest) signatures attempts, even during the TEST in the distribution step of the protocol.
Thus it can be shown \cite{pab+20} that having $p_e < s_a$ we have a probability of honest abort:
\begin{equation}
    p_{\text{honest-abort}} \leq \exp(-c_{ha}(p_e-s_a)^2 n)
,\end{equation}
with some constant $c_{ha} > 0$, thus the protocol is ``robust'' against honest aborts.

\paragraph{Choosing the Optimal Thresholds}
Following \cite{pab+20}, with the probabilities of forgery, repudiation and honest abort in mind, we find that for our qandy signature protocol the following relation must hold:
\begin{equation} \label{eq:pr_rel}
    0 \leq p_e < s_a < s_v < p_f
.\end{equation}
In order to minimize the probability of any ``bad event'' (successful forgery, successful repudiation or honest reject) happening, the parameters $s_a, s_v$ must be chosen such that the distance between each pair of probabilities in eq.~\eqref{eq:pr_rel} is equal.
Of course, one may also set the parameters differently (and not symmetrically) if not all the ``bad events'' are equally important.

\section{Alternative Signature Protocols} \label{sec:alt_sig}
Given the original constraint of requiring authenticated quantum channels in P1 of \cite{wdka15}, an alternative protocol was suggested by the authors - P2.
Then, in \cite{awka16}, another protocol was suggested that also
avoids authenticated quantum channels, which is an improved version of
\cite{wdka15}'s P2, in terms of efficiency.

In this section we first discuss the P2 protocol and compare it to P1, and then discuss \cite{awka16}'s and compare it to P2. 

\subsection{Protocol P2 of \cite{wdka15} with Qandies} \label{sec:P2}

As we already clarified, 
the protocol P2 is in a sense a ``classical'' signature scheme, based on shared secret classical keys.
The secret classical keys in P2 
are obtained using QKD techniques, 
and thus the protocol may be implemented using qandies, 
by applying qandy QKD as described in \cite{lm20, lms21}.
The protocol steps are almost identical to the steps of OTP-S in Section~\ref{subsec:otps-spec}; instead of using OTP to obtain secure classical channels, the parties use qandies to implement standard QKD (thus requiring only an authenticated classical channel), including the classical post-processing steps of TEST, Error correction (EC) and Privacy Amplification (PA).

\begin{figure}[!ht]
	\centering
	\subfloat[P1 Distribution \& Symmetrization\label{fig:p1}]{%
        \includegraphics[width=0.4\textwidth]{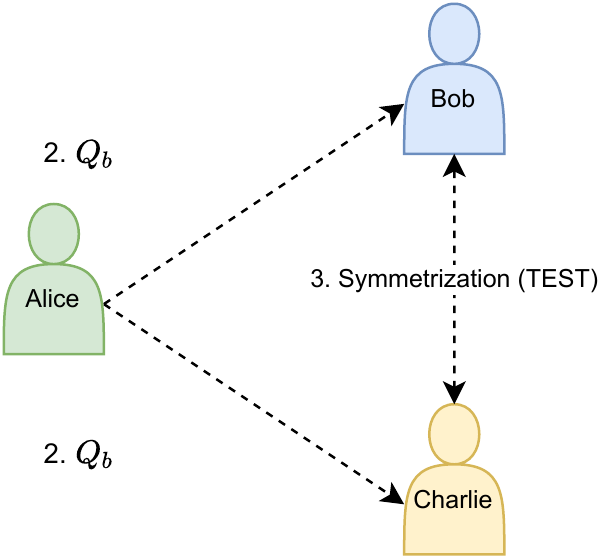}
	}
	\hspace{0.15\textwidth}
    \subfloat[P2 Distribution \& Symmetrization\label{fig:p2}]{%
        \includegraphics[width=0.4\textwidth]{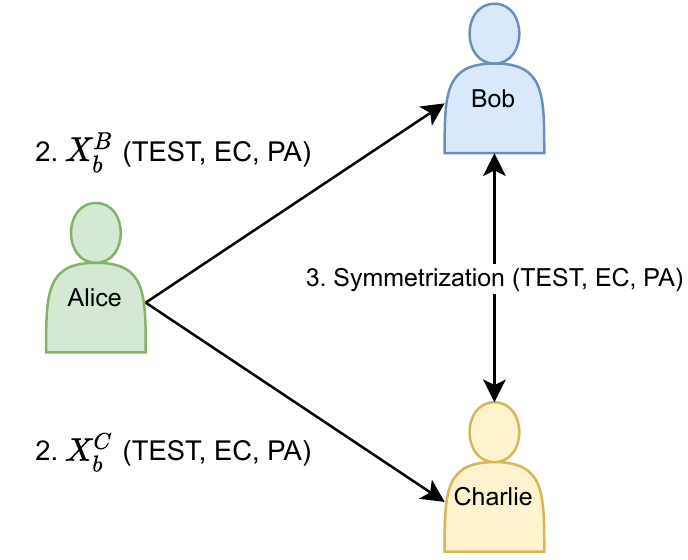}
    }
    \hspace{0.15\textwidth}
    \subfloat[AWKA Distribution \& Symmetrization\label{fig:awka}]{%
        \includegraphics[width=0.4\textwidth]{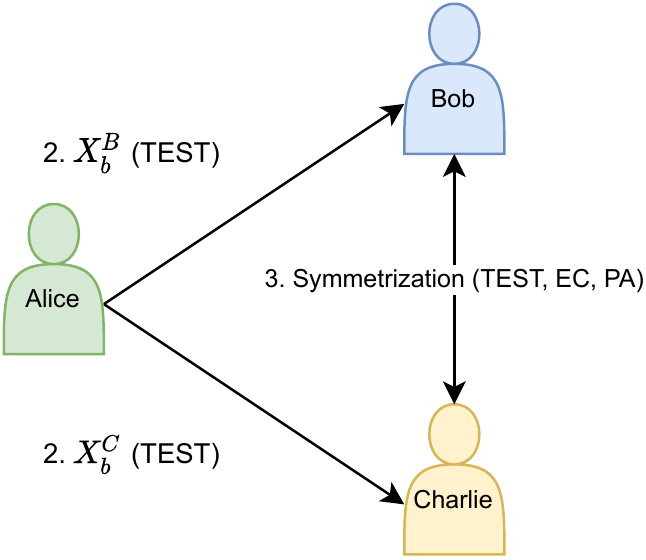}
    }
    \caption{
    ``Public'' Key Distribution (Step~\ref{step:pk_dist}) and Symmetrization (Step~\ref{step:pk_symm}) of the (a) P1, (b) P2 and (c) \cite{awka16} protocols.
    Step~\ref{step:pk_gen} (Private Key Generation, see for example Subsections~\ref{subsec:otps-spec} and \ref{subsec:p1-spec}) is not seen in the figure.
    In P1, the parties transfer qandies (dashed arrows), with TEST done after the symmetrization step.
    In P2, the parties transfer classical bits (solid arrows), achieved via QKD over insecure ``qandy channels'', with full classical post-processing (TEST, EC and PA).
    In \cite{awka16}, the protocol is similar to P2, achieved via (partial) QKD over an insecure ``qandy'' channel, except no EC and PA are done in Step 2.
    }
    \label{fig:protocols_fig}
\end{figure}

We now provide several remarks on the protocol and compare it to P1:
\paragraph{Classical Public Keys}
The main difference of this protocol from the P1 protocol of \cite{wdka15} is that it uses \emph{classical} public keys instead of quantum/qandy keys, and only the method to obtain those keys is quantum/qandy based (see Figure~\ref{fig:protocols_fig}).
The downside vs. P1 is that due to employing full QKD (and not just the TEST stage as in P1), Alice may need to send longer strings of qandies to achieve a desired level of security.
This disadvantage is later resolved by \cite{awka16}, by replacing the full QKD in the distribution step with a variant that includes just messaging + TEST, essentially shifting the protocol back towards P1 in this sense (see Section~\ref{subsec:awka16}).
\paragraph{Usage of Memory}
In the symmetrization step of the original P2 protocol by \cite{wdka15}, Bob and Charlie randomly decide whether to forward each bit of the public key to the other, and abort the protocol if one of them received from the other a number of bits too far from $n/2$.
Here we decided to have Bob and Charlie randomly send \emph{exactly} half the bits to the other, like in our description of P1, in order to simplify the description.
Contrary to P1, having Bob and Charlie store their bits in memory here is not difficult to implement, as classical memory is usually a trivial resource.
\paragraph{Verification Threshold}
In the verification step, note that in contrast to P1, we have no threshold $s_a$ and Bob must ensure that \emph{all} the bits he has in his possession (both from $X^B_b$ and $X^C_b$) match Alice's declaration.
This is because the public keys were distributed with no errors at all over the secure classical channels during the distribution step.

\paragraph{Protocol Security}
As the OTP-S protocol is proven secure by \cite{wdka15}, and QKD with qandies is secure \cite{lm20, lms21}, the P2 protocol is also secure.

\subsection{Protocol of \cite{awka16} with Qandies} \label{subsec:awka16}
Following the work of \cite{wdka15}, an improved version of P2 was proposed and proven secure in \cite{awka16}.
\cite{awka16} is a somewhat ``intermediate'' protocol between \cite{wdka15}'s P1 and P2 protocols.

Similarly to P2, the protocol uses shared classical keys between the participants, with Bob and Charlie each getting a different key.
The difference from P2 is that instead of employing all stages of QKD to obtain fully identical and fully secret keys, \cite{awka16} avoid implementing the EC and PA stages during the key distribution step, i.e. only implement the TEST stage, as in P1 (see Figure~\ref{fig:protocols_fig}).
This allows Alice to send fewer qubits in order to obtain a desired level of security (since no bits are ``sacrificed'' for EC and PA).

Due to skipping EC and PA, the generated keys are not entirely identical and not entirely secret, but are highly correlated (in fact, as showed by \cite{awka16}, more correlated than with any key an eavesdropper could produce).
This enables the signature protocol (which is identical to that of P2 in the rest of the steps) to remain secure, although the security proof is much more complicated in this case, and can be found in \cite{awka16}.
We note that in their security proof, \cite{awka16} assume that it is Bob who is the sender in the QKD of the distribution step, to simplify the security analysis of non-repudiation.
Regardless, according to \cite{awka16} it is not mandatory and the protocol may still work even when Alice is the sender.

Of course, the same protocol may be implemented using qandies - we implement qandy QKD to share the keys between Alice and Bob/Charlie, without the post-processing steps of EC and PA.
The rest of the protocol follows analogously.

\section{Conclusion} \label{sec:conclusion}

In this paper we have discussed the topic of digital signatures and the qandy
model proposed by Jacobs (presented and extended in \cite{lm20,lms21}).

Our main idea in the current paper was using 
the qandy model to present 
a simplified version of the P1 quantum digital signature protocol 
presented by \cite{wdka15}.
Then, we showed that the protocols P2 of \cite{wdka15} and the protocol of \cite{awka16} may be translated to the qandy world as well.
We also clearly presented a fully classical method (a classical variant
of P2) that is due to \cite{wdka15}, yet its connection
to OTP is somewhat hidden in the original paper, so we thought it is important
to highlight it here.
 
The work is meant to serve mostly as a continuation to the ideas presented in
\cite{lm20,lms21}, with the goal of presenting interesting concepts from quantum information science to the unfamiliar reader, focusing on quantum digital signatures.
While not many quantum-information protocols can be defined without using superpositions, unitary transformations, and phases, we showed here it is possible for quantum digital signatures.

\appendix
\section{Appendix} \label{sec:append}

\subsection{Classical Digital Signature -- Formal Definition}
We now present a conventional definition for a correct and secure one-time digital signature scheme. 
The definition states the Syntax of the scheme, i.e., which functions are required, the meaning of a ``correct'' scheme and the meaning of a ``secure'' scheme.
\paragraph{Syntax}
A (classic) one-time
digital signature scheme $\Pi$ is defined by three probabilistic polynomial-time algorithms $\Pi$ = (Gen, Sign, Ver) with the following syntax:

\begin{itemize}
    \item Gen($1^n$)$\to (vk,sk)$
    \item Sign$(sk,m) = \sigma$
    \item Ver$(vk,m,\sigma)\to$ ACC/REJ
\end{itemize}
Where $vk$ is a public verification key, $sk$ is a private signing key and $\sigma$ is a signature for a message $m$.

\paragraph{Correctness}
A digital signature scheme is said to be \textbf{correct} if the following equality holds:

\begin{equation}
    p[\text{Gen}\to(vk,sk) ;\text{Ver}(vk,m,\text{Sign}(sk,m)) = \text{ACC}]=1
\end{equation}
I.e., for every generation of the verification key $vk$ and the signing key $sk$, the algorithm Ver will accept every signature for a message $m$ that was made with Sign for the same message $m$, with certainty.
This should hold for every message.
\paragraph{Security}
A digital security scheme is said to be \textbf{secure} if any efficient adversary $A$ wins the following game (``The security game'') with negligible\footnote{a function $\epsilon:\mathbb{N} \to \mathbb{R}$ is called negligible if for every polynomial function P(x) there exists $n_0$ such that for all $n>n_0$, $\epsilon(n)< \frac{1}{P(n)}$}
probability:
\begin{enumerate}
        \item 
        Gen$(1^n)\to (pk,sk)$
        \item 
        $A^\text{Sign(sk,$\circ$)}(1^n)\to (m^*,\sigma^*)$ 
    \end{enumerate}
    Where Sign$(sk,\circ)$ represents \textbf{one query} by $A$ to an oracle who can sign any chosen message $m$ by $A$.
    $A$ wins if Ver$(vk,m^*,\sigma^*)$ = ACC and $m^*\neq m$.
    
\paragraph{}
We should note that there are stronger definitions for security, which define different schemes.
For example a ``digital signature scheme'' (as opposed to one-time DS scheme) allows security against an adversary with polynomial number of queries to the oracle.
One-time secure digital signatures schemes may still be relevant for some application, and also can be the basis for more complex schemes with stronger notions of security.
Modern QDS protocols deal with one-time schemes for one bit messages, and this is also the type of protocol we presented in this paper using quantum candies.

\subsection{Lamport Scheme -- Intuitive and Formal Definition}

We repeat here the protocol specifications provided in Section~\ref{sec:Lamport},
this time also mentioning in parentheses the relation to the formal definition. 

\paragraph{}
Let Alice be the sender of a message $m \in \{0,1\} $, Bob the recipient and Charlie the arbiter in the case of dispute between Bob and Alice. Let $H: \{0,1\}^n\to\{0,1\}^n$ be a one-way function.

\begin{enumerate}
    \item \textbf{Private key generation}
    For each value of $b\in \{0,1\}$, Alice generates a random string $X_b\in\{0,1\}^n$.
    Both $X_b$ serve as the private keys of Alice (so $sk=(X_0,X_1)$).

    \item \textbf{Public key distribution}
    For each value of b, Alice calculates and publicly announces $P_b=H(X_b)$.
    Both $P_b$ serve as the public keys (so $vk=(H,P_0,P_1)$. These two procedures, 1 and 2, define Gen).

    \item \textbf{Signed messaging}
    If Alice wishes to send Bob a signed one-bit message $m$, she sends $(m, X_m)$ (so Sign$(sk,m)=sk[m])$). 
    I.e, the signature of the message $m$ is\footnote{Note that in this construction, $sk=[\sigma_0,\sigma_1]$} $\sigma_m=X_m$

    \item \textbf{Verification}
    When receiving $(m, X_m)$ from Alice, Bob (who knows $H$ and $P_m$) accepts the message if and only if $H(X_m)=P_m$ (This defines Ver).

    \item \textbf{Arbitration}
    A dispute occurs when Bob claims Alice sent a message $m$, and Alice denies it.
    If Bob is dishonest, this is the forgery attempt case.
    If Alice is dishonest, this is the repudiation attempt case.
    A dispute is settled in the following way: Bob sends Charlie the value of $X_m$ he claims he received from Alice, $X^B_m$.
    Charlie then checks if $H(X^B_m)=P_m$, and if so he declares Alice as the dishonest one.
    If not - he declares Bob as the dishonest one.

\end{enumerate}

\textbf{Unforgeability} is guaranteed since if $H$ is a one-way function, Bob will not be able to find such $X^B_m$ (within reasonable time).
\textbf{Non-repudiation} is guaranteed for the same reason - if Bob managed to present such $X^B_m$, then he definitely got it from Alice.

\bibliographystyle{alpha}
\bibliography{qandySig}

\end{document}